\documentclass[eqsecnum,twocolumn,showpacs,preprintnumbers,amsmath,amssymb]{revtex4}

\usepackage{graphicx}
\usepackage{dcolumn}
\usepackage{bm}
\usepackage{epsf}
\usepackage{epsfig}
\usepackage{psfig}
\usepackage{graphicx}

\begin{document}


\title{Coherent Transport and Concentration of Particles in Optical
Traps using Varying Transverse Beam Profiles}

\author{Ole Steuernagel}
 \email{ole@star.herts.ac.uk}
\affiliation{Dept. of Physics, Astronomy, and Mathematics,
University of Hertfordshire, College Lane, Hatfield, AL10 9AB, UK}%

\date{\today} 

\begin{abstract}
Tailored time-dependent variations of the transverse profile
together with longitudinal phase shifts of laser beams are
studied. It is shown theoretically that a standing wave setup and
real-time beam forming techniques (e.g. by computer-addressed
holograms) should make it possible to implement smooth transport
across and along the beams employed in optical trapping schemes.
Novel modes for the efficient collection, transport, and
concentration of trapped particles should thus become realizable
in optical trapping setups.
\end{abstract}



\maketitle

%
\section{Introduction}
%
Trapping of objects with light is possible in all transparent
media such as liquids, air and other gases, and vacuum. Laser beam
trapping has become an established technique where the size of the
trapped objects ranges over many orders of magnitude from atoms to
particles of several hundred $\mu m$
size~\cite{Nature.tweezer.review}. There are two standard
configurations: firstly counter-propagating plane waves form
standing light field patterns yielding multiple traps arranged as
crystals (i.e. periodic intensity patterns or 'light
crystals'~\cite{light.crystals}), and secondly, strongly focussed
laser beams form laser tweezers, whose foci serve as single
trapping centers~\cite{Nature.tweezer.review}. Because of their
great power concentration, laser tweezers can levitate and hold
small beads of many micrometers size. Smaller objects such as
bacteria, nano-particles, molecules and atoms can also be held and
moved by 'light crystals'. For several atomic species
magneto-optical~\cite{magneto.optical} and all-optical cooling
schemes~\cite{all.optical} have, moreover, allowed to create
ultra-cold samples of dilute gas, some of them at fraction of
nanokelvin temperatures in the Bose-Einstein-condensate
state~\cite{Leanhardt03}.

Although trapping (and cooling) particles with light is now a well
established and mature field, moving such trapped particles with
the help of the trapping fields is less refined. The main purpose
of this paper is the introduction of a new approach to the
collection, (coherent) transport, and { spatial concentration} of
particles.

In particular the spatial concentration of particles with current
schemes is not optimized: in the case of plane wave generated
light crystals the crystal cells cannot be merged, laser tweezers
suffer from small focal volumes, and optical washboard
potentials~\cite{washboard} do not transport coherently. So far,
spatial concentration towards one point is only achieved by
changes to an auxiliary potential such as that of an assisting
magnetic field~\cite{magnetic.compression}. Here, it is shown that
modulation of the beam characteristic of a laser tweezer itself
can open up new modes of coherent transport, capture,
concentration, excitation, and release of particles.

For trapped atomic clouds this can help to increase their phase
space density since they can be simultaneously cooled and
spatially concentrated~\cite{all.optical}. Likewise, larger
particles suspended in viscous media can be concentrated in phase
space. In the case of cold atomic clouds it might help to
continuously replenish lossy traps~\cite{Gustavson02} thus leading
to continuous, Bose-Einstein-condensation--mediated, atom-laser
operation~\cite{Koehl01,{Chikkatur02}}. In the case of ions,
fermions and other mutually repulsive particles their collection
and spatial concentration might make it easier to reach
unit-filling factors for the particle population trapped in an
optical lattice~\cite{unit.fill}; useful, e.g., for grid-based
quantum computing~\cite{Raussendorf01}.

This paper will first review current setups and outline my
approach in section~\ref{technology} and the terminology for the
description of paraxial beams in section~\ref{gauss.beams}.
Section~\ref{Examples} and its subsections will deal with one- and
two-dimensional modifications of the transverse mode profile of
paraxial beams over time. This will demonstrate that manipulating
the mode structure of a laser beam enables us to tailor its
structure in such a way that controlled transport of trapped
particles across the beam becomes possible and that this can be
designed such that the particles are moved into a smaller volume
by merging the cells of the effective light lattice that traps the
particles. For further concentration an optical conveyor belt is
introduced allowing us to concentrate particles towards a 'point'
in space and unload them there. After that, coherence preserving
transport is considered. Next, these ideas are generalized for the
case of low-field seeking trap particles in
section~\ref{low.field.seeker} followed by the conclusion.
%
\section{Current Setups and Possible Modifications}\label{technology}
%
For focussed beams several scenarios have been implemented:
longitudinally moving the trapping center by refocussing together
with redirection of the beam axis allows for three-dimensional
movement of the focus~\cite{Gustavson02}. But the focal volume is
quite small and has to be moved in order to efficiently pick up
more particles. Therefore multi-beam approaches in which many
independent Gaussian foci are created have been demonstrated by
holographical beam
splitting~\cite{{computer.hologram},holographic-beam-splitting-Grier},
including their independent
movement~\cite{{computer.hologram},independent.foci},
merger~\cite{peristaltic.foci}, and application for size-selective
particle deflection~\cite{fractionation.Grier}. Yet, these
methods~\cite{{computer.hologram},holographic-beam-splitting-Grier,independent.foci,peristaltic.foci}
do not continuously collect over the entire beam volume but rather
'pointwise' at the various foci.

In focussed beams also the transverse beam profiles have been
changed to generate annular high-order TEM-modes yielding 'optical
tubes'~\cite{optical.tubetrap,{optical.tubes}}, 'optical
bottles'~\cite{optical-bottles} and beam centers surrounded by
washboard potentials~\cite{washboard}. Moreover, beams have been
equipped with orbital angular momentum thus allowing trapped
particles to be turned~\cite{turn.light} and, also, to use this
freedom for quantum information coding~\cite{qudit}. Finally,
tilted reflecting light sheets together with gravity have been
used to implement atomic billiards~\cite{opt.billiard}.

For counter-propagating plane waves (wide laser beams) light
crystals with different symmetries are real\-iz\-able. Depending
on the number and relative orientation of the employed beams, they
form, e.g., cubic, tetrahedral, and
super-lattices~\cite{superlattice} in three dimensions.
Effectively two-dimensional sheets formed by evanescent
waves~\cite{gost} and arrays of one-dimensional
tubes~\cite{Greiner01} have also been implemented. Such crystals
can be moved by detuning the frequency~\cite{accel.lattice} or
otherwise shifting the relative phase between beams, their lattice
constants can be varied to some extent by changing the relative
angles between interfering beams~\cite{light.crystals}, but their
unit cells cannot be merged. I therefore want to explore other
avenues for the transport and concentration of trapped particles.
%
%

Currently established setups are quite static in the sense that
the underlying optical beam shapes are kept
unaltered~\cite{McGloin03}. The following changes to the beam
structure could be considered though: one could change the
longitudinal properties of the beam, but for beams propagating
freely in homogenous media this is typically done by changes to
their spectral composition, for instance by
frequency-sweeping~\cite{accel.lattice} or pulsing the
beam~\cite{Kuhr01}. Here, only gradual and slow changes which, in
the case of quantum particles, will also allow us to preserve the
trapped objects' coherence~\cite{{Haensel01},Orzel} are
considered. Also, the polarization state of the trapping light
fields could be changed over time, this will be further
investigated in future work. Instead, the present paper
concentrates on slow temporal variations to the beam's transverse
field and intensity profiles for uniformly focussed polarized
light beams~\cite{patent}.
%
%

\begin{figure}
\epsfverbosetrue \epsfxsize=3.4in \epsfysize=1.5in
\epsffile[030 050 720 620]{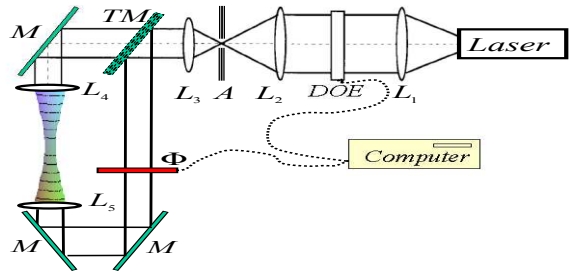}
\caption{Sketch of possible technical implementation (not to
scale): a setup using laser, five lenses and aperture $A$, a
semi-transparent balanced mirror $TM$ to split the beam and three
mirrors $M$ to recombine the two parts into one
counter-propagating standing wave beam. A computer-controlled
diffractive optical element $DOE$ generates the modulation of the
trapping beam (which is shown as a solid hyperboloid). A
phase-shifter~$\Phi$ allows for computer-controlled longitudinal
shift of the beam. \label{setup.trombone.phase}}
\end{figure}

The field of diffractive optics is a mature field routinely using
computer-generated holograms to alter light fields. Liquid crystal
arrays and other spatial light modulators, have been developed for
video-beamer technology but are now also used to implement
diffractive optical elements with computer-generated holograms in
real
time~\cite{Nature.tweezer.review,computer.hologram,holographic-beam-splitting-Grier,independent.foci,peristaltic.foci,CGH,{cgh.Liesener}}.
Typically, the diffractive element is positioned in a region where
the laser beam is wide and its wave fronts parallel, see
Figure~\ref{setup.trombone.phase}. It imprints its amplitude
information on the wide parallel beam the width of which then is
suitably shrunk (using lenses $L_2$ and $L_3$ in
Figure~\ref{setup.trombone.phase}). Finally the beam is collimated
using another another set of lenses ($L_4$ and $L_5$ in
Figure~\ref{setup.trombone.phase}). Since the focus of the
collimated beam is --up to rescaling and redirection-- the
Fourier-transform of the diffractive element's amplitude pattern,
we can easily calculate the required input with a computer that
controls the diffractive optical element in real time.

Resolution of the diffractive optical elements are not a problem
since mega-pixel LCD-screens are commercially available. Also the
deviations of the diffractive element's input from the ideal
pattern due to its pixelated structure is not a problem. The
regular pixelation gives rise to diffraction off axis which can be
filtered out using an aperture $A$ serving as an effective
low-pass filter~\cite{cgh.Liesener}.

If the phase shift $\Phi(t)$ is implemented by shifting the
frequency of one of the counter-propagating beams with respect to
the other, very large phase differences can be accumulated very
quickly. The same is currently not yet true for modifications of
$I(x,y,z;t)$, beam formers were developed for video technology and
only allow to modify the transverse intensity profile at video
frame rates, i.e. on the order of some hundred
Hertz~\cite{Nature.tweezer.review,computer.hologram,holographic-beam-splitting-Grier,independent.foci,peristaltic.foci,CGH,{cgh.Liesener}}.
Fortunately, this is not a fundamental limit and it should be
easily overcome in the near future~\cite{McGloin03}.
%
%
%
\section{Hermite-Gaussian Beams: TEM-modes}\label{gauss.beams}
%
In practical applications laser beams which are not too tightly
focussed are very important. Although the ideas presented here are
in principle applicable in more general cases, e.g. for very
tightly focussed beams or very general fields created by intensity
masks or holograms, we will only consider quasi-monochromatic
beams in the paraxial scalar
approximation~\cite{Yariv.buch,Haus.buch}. In this approximation
the solutions are the familiar {\em transverse electro-magnetic}
or TEM$_{mn}$ modes describing $x$-polarized beams propagating in
the $z$-direction with a vector potential ${\bf A} = (A_x, A_y,
A_z)$ whose only non-zero component is $A_x$ with~\cite{Haus.buch}
%
\begin{equation}
A_x({\bf r},t;k) = \psi_{mn}({\bf r})\; e^{i(k z - \omega t)},
\label{A}
\end{equation}
%
where the scalar function~$\psi_{mn}$ contains products of
Gaussians and Hermite-polynomials, i.e. the familiar harmonic
oscillator wave functions~~$\varphi_m(\xi) =
H_m(\xi)\,\exp(-\xi^2/2) / \sqrt{2^m m! \sqrt{\pi} } $,
$(m=0,1,2,\ldots),$ and various phase factors~\cite{Haus.buch,{Yariv.buch}}
%
\begin{eqnarray}
\psi_{mn}({\bf r}) =\frac{w_0}{w(z)} \;
\varphi_m(\frac{\sqrt{2}\,x}{w(z)})\;
\varphi_n(\frac{\sqrt{2}\,y}{w(z)})
\\ \nonumber
\times \; e^{\frac{i k}{2 R(z)}(x^2+y^2)} \; e^{-i (m+n+1)
\phi(z)} \; . \label{HG.TEM}
\end{eqnarray}
%
The dispersion-relation of light in a homogenous medium $\omega =
c k$ was used; $x,y$ are the transverse and $z$ the longitudinal
beam coordinate, $t$ is time and $w_0 = \sqrt{2b/k} =
\sqrt{\lambda b/\pi}$ is the relation that links the minimal beam
dia\-meter $w_0$ with the Rayleigh range~$b$. The beam dia\-meter
at distance $z$ from the beam waist $(z=0)$ obeys $w(z)=
\sqrt{w_0^2 (1 + z^2/b^2)}$ and for large~$z$ shows the expected
amplitude decay of a free wave~$\propto 1/|z|$, the beam's opening
angle in the far-field is $\arctan(\lambda/(\pi w_0))$. The
corresponding wave front curvature is described by $R(z)= (z^2 +
b^2)/z$, and the longitudinal phase shift (Gouy-phase) follows
$\phi(z)=\arctan(z/b)$; according to the Gouy-phase factor $e^{-i
(m+n+1) \phi(z)}$ it leads to relative dephasing between different
modes.

The vector potential $A_x$ of Equation~(\ref{A}) describing a beam
travelling in the positive $z$-direction (${\bf k}=k\hat{\bf z}$)
yields an electric field which is polarized in the $x$-direction
with a small contribution in the $z$-direction due to the tilt of
wave fronts off the beam axis ($\hat{\bf x}, \hat{\bf y}, \hat{\bf
z}$ are the unit-vectors). We omit this wavefront tilt and hence
only deal with the scalar approximation
%
\begin{eqnarray}
{\bf E} \approx E_x \; \hat{\bf x} = \Re \left\{  \omega A_x \;
\hat{\bf x} \right\} \; .
\label{scalar.E.field}
\end{eqnarray}
%
Just like the paraxial approximation, the scalar approximation
gets better the less focussed the beam (the larger the beam waist
$w_0$) is.

Since the wave equation is linear and the harmonic oscillator wave
functions form a complete orthonormal set for the transverse
coordinates $x$ and $y$, we are free to combine the above
solutions to generate many interesting field and intensity
configurations~\cite{Rayleigh.limit}
%
\begin{eqnarray}
A_x({\bf r},t;{\bf k}) = \sum_{m,n=0}^\infty  c_{mn}(t) \;
\psi_{mn}({\bf r})
 \; e^{i(k z - \omega t)} . \label{sum.TEM}
\end{eqnarray}
%
The coefficients $c_{mn}(t)$ can be complex (i.e. change amplitude
and phase of the beam), can be varied with time and do not obey
normalization restrictions. Since we have trapping in mind let us
also assume that we discuss standing wave fields formed from a
superposition of (other\-wise identical) counter-propagating
beams, see Figure~\ref{setup.trombone.phase} above. In this case
we have
%
\begin{eqnarray}
A_x = \sum_{m,n=0}^\infty  c_{mn}(t) \; \psi_{mn}({\bf r})
 \; e^{i (kz - \omega t +\Phi(t)) } \; + c.c.,
\label{A.standing.wave.sum.TEM}
\end{eqnarray}
%
where $\Phi(t)$ represents the controllable, relative phase shift
between the two beams forming the standing wave pattern and $c.c.$
stands for complex conjugate. The resulting intensity distribution
$I(x,y,z;t)\propto {\bf E}(x,y,z;t)^2$ only contains terms with a
controllable (slow) time-dependence, namely $c_{mn}(t) $ and
$\Phi(t)$ (see remarks at end of section~\ref{technology}).
%
\section{Examples}\label{Examples}
\subsection{Transverse 2-D Profiles}\label{2D}
%
%
\begin{figure}
\epsfverbosetrue \epsfxsize=3.4in \epsfysize=2.4in
\epsffile[10 120 500 605]{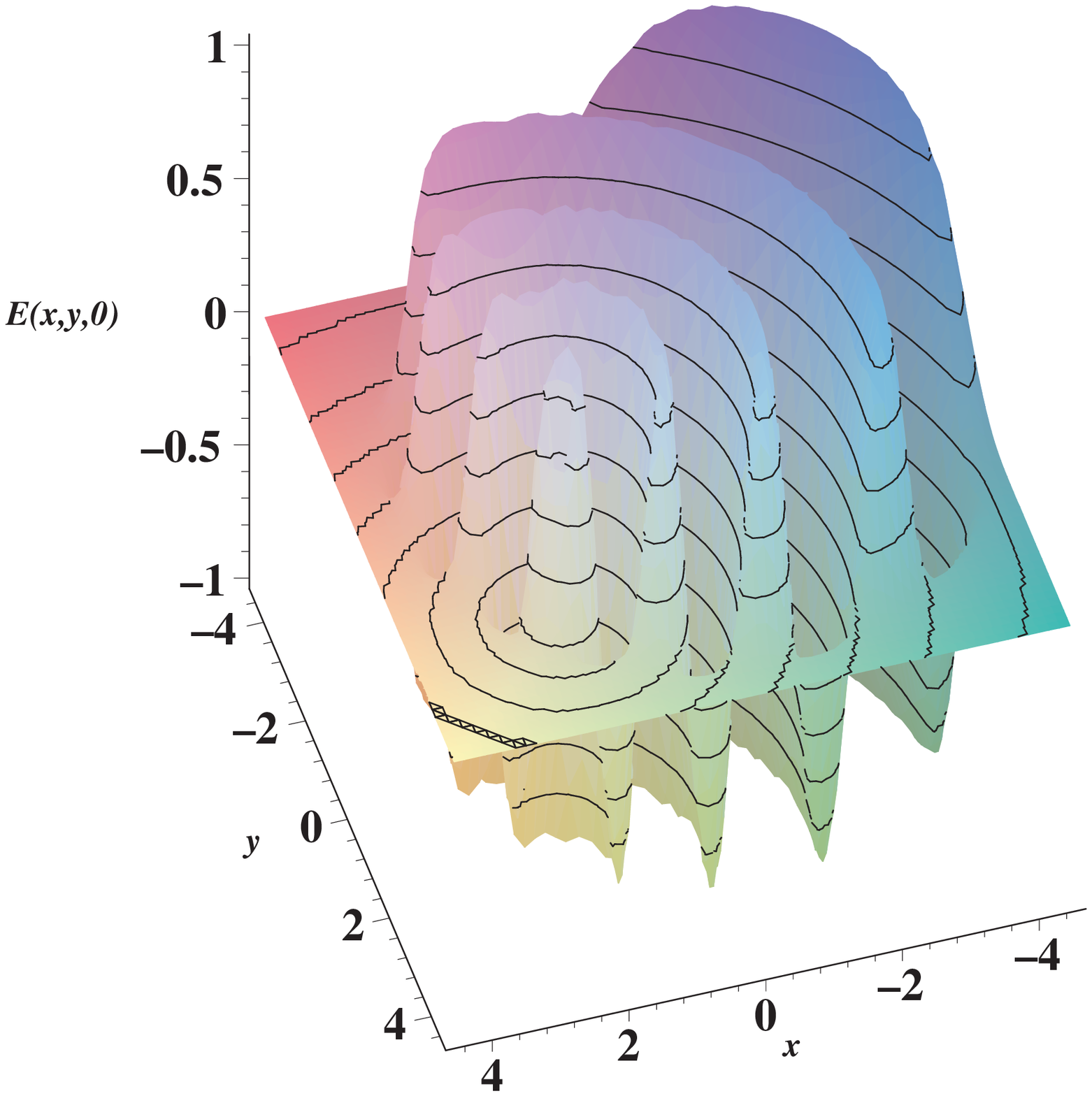}
\caption{Plot of a possible transverse field configuration at
$z=0$, i.e. a slice across the laser beam. The field forms ring
structures converging at position $(x,y)=(2,2)$.
\label{concentric.E}}
\end{figure}
%
As an example Figure~\ref{concentric.E} specifies a possible field
configuration of concentric waves emerging at the periphery of the
trapping beam which then travel across the beam converging at one
point [at $(x,y)=(2,2)$] on the opposite edge thus concentrating
all captured particles into a perl string, on the beam's fringe,
such as that depicted in Figure~\ref{4_I5} below.
Figure~\ref{c.nm7} displays the expansion coefficients $c_{mn}(t)$
at one particular moment in time $t$ up to twelfth order in $m$
and $n$ and Figure~\ref{c.n3T} depicts the time-development of a
subset of the coefficients~$c_{mn}(t)$ and displays the periodic
motion underlying the concentration process portrayed above.

%
\begin{figure}
\epsfverbosetrue \epsfxsize=3.4in \epsfysize=2.4in
\epsffile[020 120 490 570]{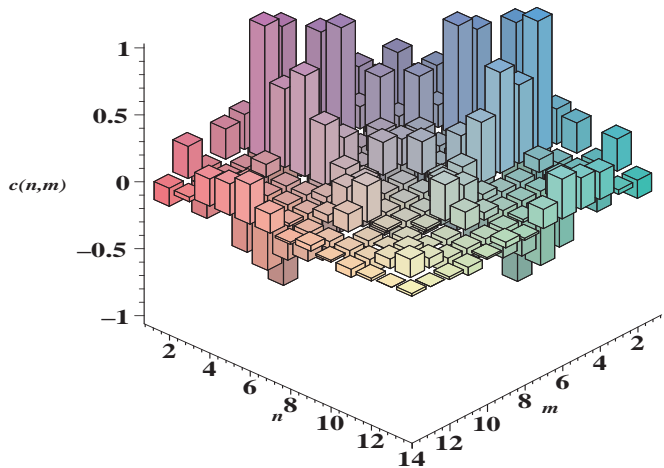}
\caption{Expansion coefficients $c_{nm}$ up to 12-th order
$n,m=0,...,12$ for the field shown in Figure~\ref{concentric.E}.
The coefficients are real numbers because the electric field is
chosen to be real, the exchange-symmetry of the coefficient
$(n\leftrightarrow m)$ is due to the field's symmetry:
$E(x,y)=E(y,x)$. \label{c.nm7}}
\end{figure}
%
%
\begin{figure}
\epsfverbosetrue \epsfxsize=3.4in \epsfysize=2.4in
\epsffile[0 110 500 590]{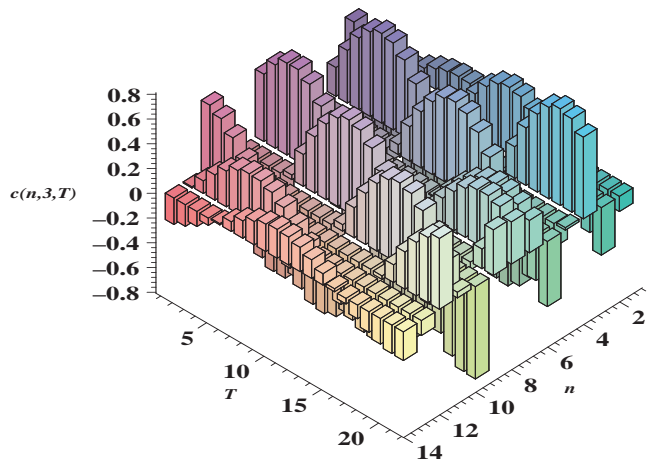}
\caption{Time-variation of a subset of the coefficients displayed
in Figure~\ref{c.nm7}. With this kind of (imposed) sinusoidal
time-dependence the intensity pattern displayed in
Figure~\ref{I.xyZT} smoothly and periodically converges towards
the concentration point on the beam edge.
\label{c.n3T}}
\end{figure}
\begin{figure}
\epsfverbosetrue \epsfxsize=3.4in \epsfysize=2.0in
\epsffile[70 90 490 660]{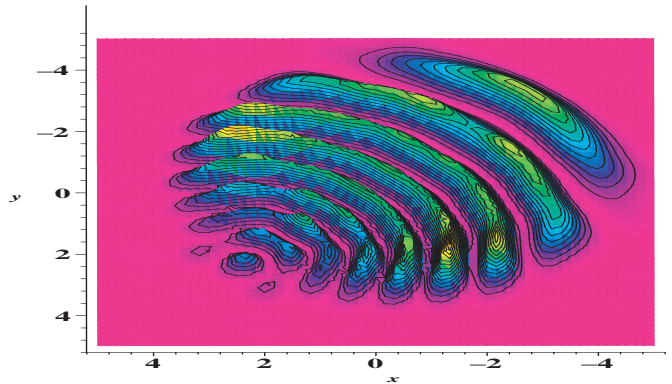}
\caption{Plot of the intensity distribution $I(x,y,0;0)$
associated with the transverse field displayed in
Figure~\ref{concentric.E} where the transverse field modes have
been determined up to 12th order: compare Figures~\ref{c.nm7}
and~\ref{c.n3T}. \label{I.xyZT}}
\end{figure}

%
Figure~\ref{I.xyZT} shows the intensity corresponding to the field
configuration displayed in Figure~\ref{concentric.E} reconstructed
using the expansion coefficients of Figure~\ref{c.nm7}. In this
case expansion up to twelfth order gives satisfactory results.
%
%
This allows us to create tailored two-dimensional potential
landscapes that can be changed over time. In particular particle
concentration, tunnelling or classical escape scenarios could be
implemented in this way~\cite{Tunneling,driven.devices}.

Once the field is specified in this way at one beam plane this
constitutes initial conditions which determine the shape along the
rest of the beam. The analysis of the resulting overall beam
behaviour and its possible applications are our next topics.
%
\subsection{3-D Concentration `in a Point'}\label{3D}
%
Let us first consider some motivation for the following
considerations: let us assume that we try to optically manipulate
particles, we want to coherently transport, concentrate and,
finally, release particles. For the particle release into a small
volume we want to assume that there is some kind of
background-trap into which we want to unload particles. We imagine
that we have captured, concentrated, and transported them using a
'foreground'-trap which relies on the methods described above.

The background-trap's field must be sufficiently strong to hold
particles but weak compared with the foreground field. Such a
background-trap could be a single laser tweezer focus, a magnetic
trap, or it could form a light crystal. We will see that even in
the case of light crystals, with their rather uniform trapping
power it is possible for the foreground beam to dominate the
particles' behavior throughout the transport and yet release the
particles into a small area. This is achieved by an optical {\em
conveyor belt} with a well defined end.

\begin{figure}
\epsfverbosetrue \epsfxsize=3.4in \epsfysize=2.3in
\epsffile{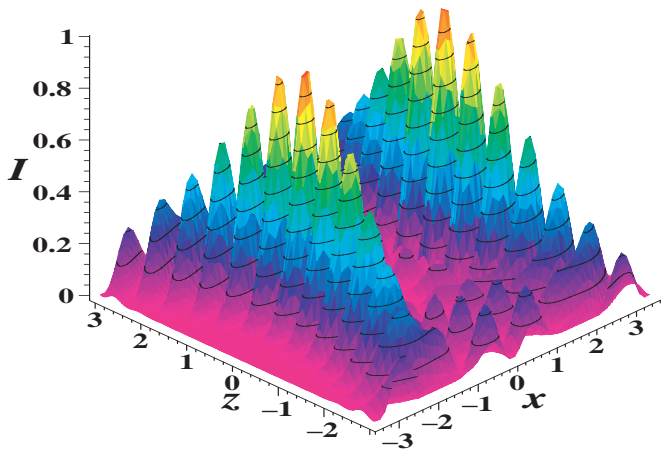}
\caption{Plot of the standing wave intensity distribution
$I(x,0,z)$ (arbitrary units) of a beam with parameter $b=3$ and
TEM-mode structure $(\varphi_3(\sqrt{2}x/w(z)) +
\varphi_5(\sqrt{2}x/w(z))) \cdot \varphi_0(\sqrt{2}y/w(z))$ near
the beam focus~$z=0$. \label{3_5}}
\end{figure}

In section~\ref{2D} we have just studied the transverse variation
of the trapping beam structure which allows us to capture
particles throughout the beam volume and within every transverse
slice concentrate particles at the beam edge. Next, we assume that
this concentration processes ceases and instead, we keep the
particles we have concentrated fixed at the beam edge. Now, by
changing the relative phase between the two beams that form the
standing wave pattern~(\ref{A.standing.wave.sum.TEM}), we can
shift the entire structure. Let us concentrate on the side which
moves towards the focus. Clearly, the foreground-trap's strength
increases near the focus. This is illustrated by the increase in
focal intensity displayed in Figure~\ref{3_5}, and nothing much is
won: if the foreground-trap manages to dominate the
background-trap elsewhere, it will typically increase its
dominance near the focus.

But, in the case of an intensity distribution which, unlike that
of Figure~\ref{3_5}, is asymmetric with respect to the beam axis,
we expect according to ray-optics that the intensity is mapped
through the beam focus; this intensity mapping can be exploited.
%
\subsection{Gouy's phase flips the intensity at the focus: the optical
conveyor belt}\label{1D}
%
%
%

\begin{figure}
\epsfverbosetrue \epsfxsize=3.4in \epsfysize=2.5in
\epsffile[35 210 590 645]{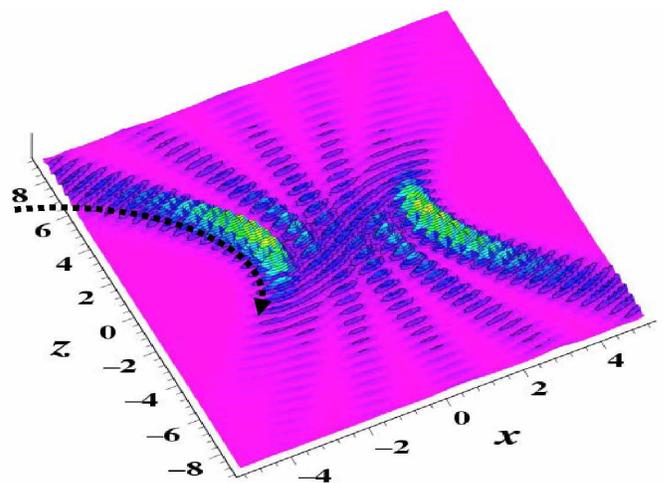}
\caption{Same as Figure~\ref{3_5} but with TEM-mode structure
$(\varphi_4(\sqrt{2}x/w(z)) + i \, \varphi_5(\sqrt{2}x/w(z)))
\cdot \varphi_0(\sqrt{2}y/w(z))$. The arrowhead indicates the
unloading point if this field configuration is used as an optical
conveyor belt. \label{4_I5}}
\end{figure}

For illustration consider the effectively one-dimen\-sio\-nal
super\-position $(\varphi_4(\sqrt{2}x/w(z)) + i
\varphi_5(\sqrt{2}x/w(z))) \cdot \varphi_0(\sqrt{2}y/w(z))$. Its
focal, standing-wave, intensity profile in the $(x,z)$-plane
$I(x,0,z)$ is shown in Fig.~\ref{4_I5} and we see that the
intensity pattern flips over when the beam passes the focal area.
%
%
It does not uniformly weaken on one side for the light intensity
to smoothly move over to the other side, instead, there is an
interesting interference scenario at the focus by which the
intensity ridge is effectively terminated on the beam edge and
separated by interference nodes from the section across the beam
where the same edge is 'resurrected' on the other
side~\cite{patent}.

If we now assume that the relative phase $\Phi(t)$ in of the
standing waves~(\ref{A.standing.wave.sum.TEM}) is varied, we
immediately see, that this can constitute a perl string of traps
which is moved towards the focus where it suddenly weakens. At the
point where the background trap is as strong as this diminishing
perl string, it starts to take over from the foreground-trap, this
combination thus forms an optical conveyor belt with a well
defined exit point where the foreground beam's cargo is handed
over to the background-trap.

\begin{figure}
\epsfverbosetrue \epsfxsize=3.4in \epsfysize=2.0in
\epsffile[100 280 850 810]{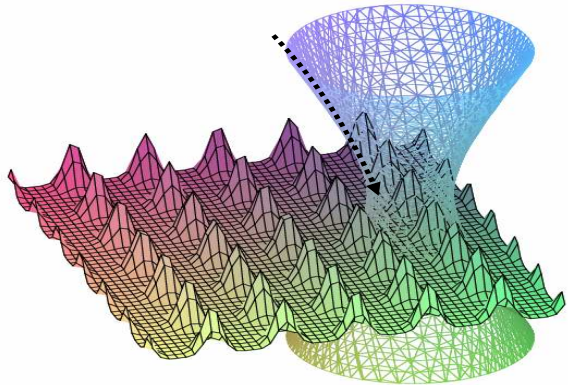}
\caption{Sketch of a possible setup: the foreground-trap shaped as
a beam funnel, feeds the background-trap, an (effectively
two-dimensional) array of laser traps formed by evanescent waves.
\label{funnel}}
\end{figure}

Fig.~\ref{funnel} sketches this scenario for the case of an
evanescent-wave light crystal that acts as a stationary
background-trapping field fed from above by the funnel-shaped
foreground field.
%
%
%
\subsection{Coherence-preserving Transport\label{coh.trans}}
%
Many examples of coherence-preserving transport of quantum
particles~\cite{Gustavson02,{Orzel}}, their
tunnelling~\cite{Tunneling,{accel.lattice}} and classical escape
dynamics~\cite{fractionation.Grier} have already been observed for
optically trapped particles. With the greater variety of trapping
potentials becoming available through the methods sketched here,
it will be possible to implement new tailored potential and thus
study such systems further.

Note, that tunnelling and classical escape processes depend
extremely sensitively (exponentially) on the potential barrier
size (Gamov-effect)~\cite{driven.devices}. In this context it is
worth mentioning that we can change the intensity ratio between
foreground and background-trap field thereby modifying the barrier
between them to make use of this exponential sensitivity. This
allows us to fine-tune the transfer process from one to the other.
%
\section{Low-field seekers\label{low.field.seeker}}
%
The discussion in section~\ref{3D} only applies to high-field
seeking particles, but for some tasks we will want to trap
low-field seekers~\cite{optical.tubetrap,optical-bottles}.
Our above discussion can be
extended to serve the case of low-field seeking particles as well,
using an altered field configuration providing us with an {\em
optical 'bubble'} or {\em 'foam' beam}.

\begin{figure}[t h]
\epsfverbosetrue \epsfxsize=3.4in \epsfysize=2.4in
\epsffile[030 50 720 570]{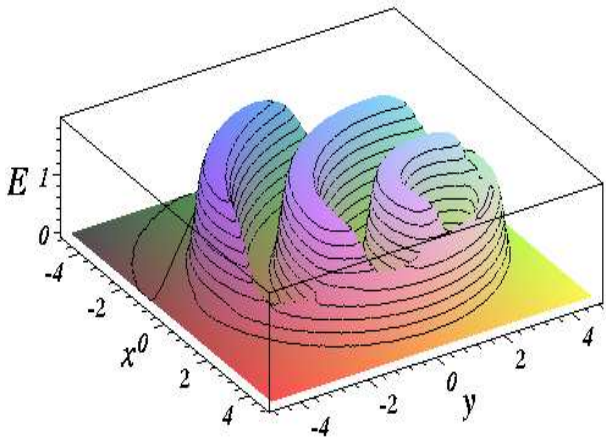}
\caption{Sketch of a possible field configuration for low field
seeking particles: the field surrounds areas of low intensity with
high intensity regions thus trapping particles in light bubbles
(and concentrating them towards the area around $(x,y)=(2,2)$).
\label{foam.field}}
\end{figure}

\begin{figure}[t h]
\epsfverbosetrue \epsfxsize=3.4in \epsfysize=1.1in
\epsffile[030 0 720 500]{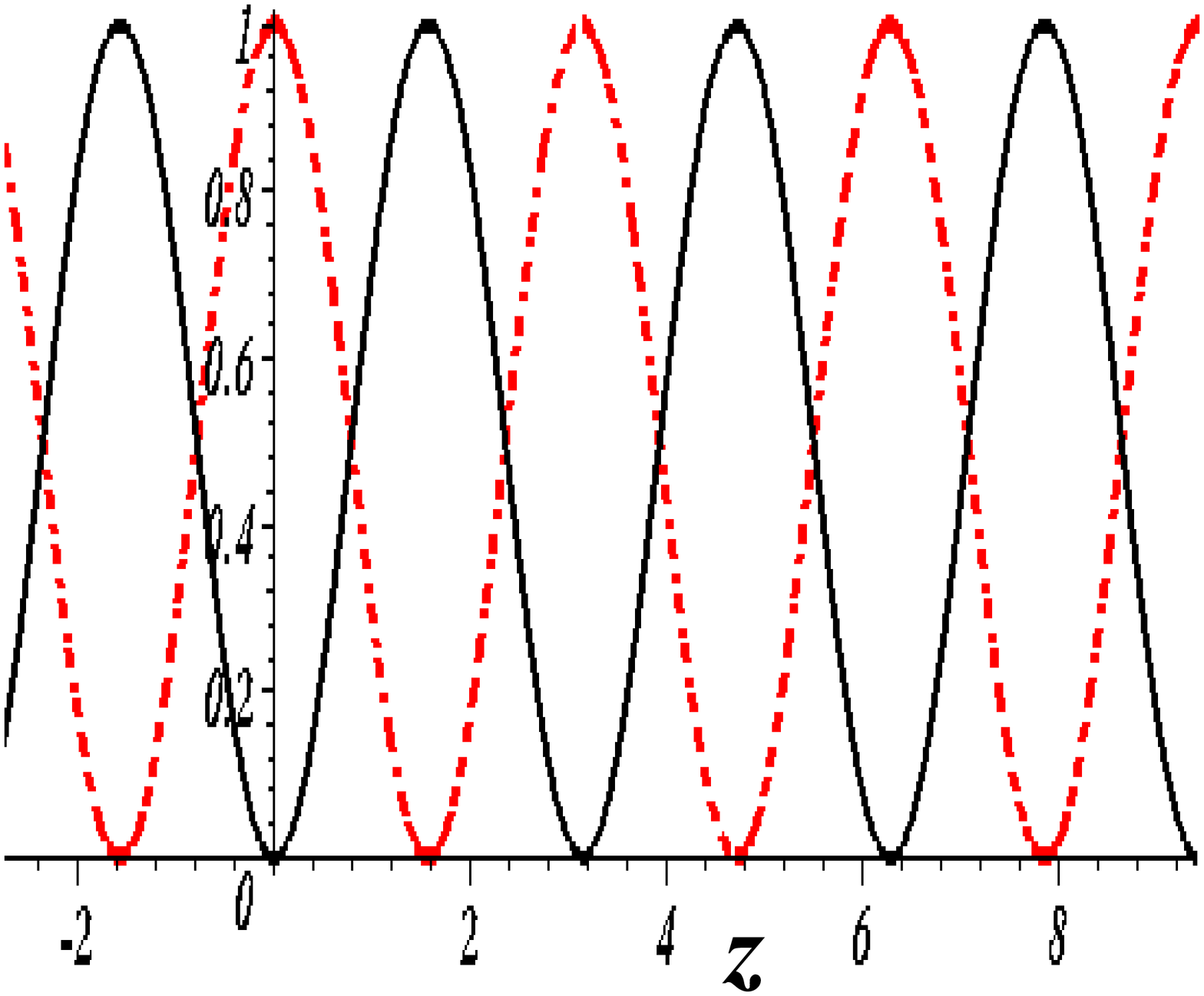}
\caption{Sketch, along the beam axis $z$, of the configuration of
the transversely modulated field (red dotted line) in conjunction
with a $90^o$-phase shifted orthogonally polarized field of equal
strength (black line) that puts an intensity plug at the nodes of
the former. According to $\sin^2 +\cos^2=1$ two waves of equal
intensity can securely encase trapped low-field seeking
particles.\label{plug.field}}
\end{figure}

As a first step, the intensity profile discussed in
section~\ref{2D} would have to be surrounded by a light rim
sealing off the beam edge and a modification of the beam such that
it contains suitable dark areas which can house low-field seeking
particles, see Fig~\ref{foam.field}. The beam would remain leaky
though, since particles could escape through the nodes of the
beams longitudinal standing wave pattern. In order to plug this
escape route one can create a second standing wave beam acting as
a stop-gap that is uniformly bright in the transverse plane and
aligned with the rest of the trapping beam, but longitudinally
shifted by a quarter wavelength. In order to avoid possible
destructive interference between these two parts of the trapping
beam they should be orthogonally polarized leading to a simple
adding up of their respective intensities, see
Fig~\ref{plug.field}. This way we can create a beam with dark
inclusions surrounded by bright areas -- a {\em 'light-foam'} or
{\em 'bubble' beam}.
%
%
%
%
%
\section{Conclusions \label{outlook.conclusion}}
%
It was shown how to implement arbitrary transverse fields with
arbitrary time-dependence, useful for trapping, coherent
manipulation, concentration and release of particles. In
particular a scheme for an optical conveyor belt with an end (in
the focal region) is introduced. Use of another interlacing trap
with orthogonal polarization was introduced in order to explain
how the ideas discussed here can be generalized to low-field
seeking trapped particles.
\begin{acknowledgments}
I wish to thank Paul Kaye, Joseph U{\l}anowski, and Ed Hinds for lively
discussions.
\end{acknowledgments}


\begin{thebibliography}{99}

\bibitem{Nature.tweezer.review} D. G. Grier,
{\em A revolution in optical manipulation}, Nature {\bf 424}, 810
(2003).

\bibitem{light.crystals} M. Weidem\"uller, A. G\"orlitz,
T. W. H\"ansch, and A. Hemmerich, {\em Local and global properties
of light-bound atomic lattices investigated by Bragg diffraction},
Phys. Rev. A {\bf 58}, 4647 (1998).

\bibitem{magneto.optical} S. Chu,
{\em The manipulation of neutral particles}; Rev. Mod. Phys. {\bf
70}, 685 (1998); C. N. Cohen-Tannoudji, {\em Manipulating atoms
with photons}, {\em ibid.} {\bf 70}, 707 (1998); W. D. Phillips
{\em Laser cooling and trapping of neutral atoms}, {\em ibid.}
{\bf 70}, 721 (1998).


\bibitem{all.optical} M. D. Barrett, J. A. Sauer, and M. S. Chapman,
{\em All-Optical Formation of an Atomic Bose-Einstein Condensate},
Phys. Rev. Lett. {\bf 87}, 010404 (2001); G. Cennini, G. Ritt, C.
Geckeler, and M. Weitz, {\em All-Optical Realization of an Atom
Laser}, {\em ibid.}, {\bf 91}, 240408 (2003).

\bibitem{Leanhardt03} A. E. Leanhardt, T. A. Pasquini, M. Saba,
A. Schirotzek, Y. Shin, D. Kielpinski, D. E. Pritchard, and W. Ketterle,
{\em Cooling Bose-Einstein Condensates Below 500 Picokelvin},
Science {\bf 301}, 1513 (2003).

\bibitem{washboard} S. A. Tatarkova, W. Sibbett, and K. Dholakia,
{\em Brownian Particle in an Optical Potential of the Washboard
Type}, Phys. Rev. Lett. {\bf 91}, 038101 (2003).

\bibitem{magnetic.compression} G. Raithel, G. Birkl, W. D. Phillips,
and S. L. Rolston, {\em Compression and Parametric Driving of
Atoms in Optical Lattices}, Phys. Rev. Lett. {\bf 78}, 2928
(1997); V. Vuletic, T. Fischer, M. Praeger, T. W. H\"ansch, and C.
Zimmermann, {\em Microscopic Magnetic Quadrupole Trap for Neutral
Atoms with Extreme Adiabatic Compression}, {\em ibid.} {\bf 80},
1634 (1998).

\bibitem{Gustavson02} T. L. Gustavson, A. P. Chikkatur,
 A.E. Leanhardt, A. Görlitz, S. Gupta, D. E. Pritchard, and
 W. Ketterle, {\em Transport of Bose-Einstein Condensates with Optical
 Tweezers}, Phys. Rev. Lett. {\bf 88}, 020401 (2002).

\bibitem{Koehl01} D. M. Stamper-Kurn, H.-J. Miesner, A. P. Chikkatur,
S. Inouye, J. Stenger, and W. Ketterle, {\em Reversible Formation
of a Bose-Einstein Condensate}, Phys. Rev. Lett. 81, 2194 (1998);
I. Bloch, M. K\"ohl, M. Greiner, T. W. H\"ansch, and T. Esslinger,
{\em Optics with an Atom Laser Beam}, {\em ibid.} {\bf 87}, 030401
(2001); M. K\"ohl, T. W. H\"ansch, and T. Esslinger, {\em
Measuring the Temporal Coherence of an Atom Laser Beam}, {\em
ibid.} {\bf 87}, 160404 (2001)

\bibitem{Chikkatur02} A. P. Chikkatur, Y. Shin, A. E. Leanhardt,
D. Kielpinski, E. Tsikata, T. L. Gustavson, D. E. Pritchard, W.
Ketterle, {\em A Continuous Source of Bose-Einstein Condensed
Atoms}, Science {\bf 296}, 2193 (2002)

\bibitem{unit.fill}  M. T. DePue, C. McCormick, S. L. Winoto,
S. Oliver, and D. S. Weiss,
{\em Unity Occupation of Sites in a 3D Optical Lattice}, Phys.
Rev. Lett. {\bf 82}, 2262 (1999).

\bibitem{Raussendorf01}  R. Raussendorf and H.-J. Briegel,
{\em A One-Way Quantum Computer}, Phys. Rev. Lett. {\bf 86}, 5188
(2001).

\bibitem{computer.hologram} M. Reicherter, T. Haist, E.U. Wagemann,
H.J. Tiziani, {\em Optical particle trapping with
computer-generated holograms written on a liquid-crystal display},
Opt. Lett. {\bf 24}, 608 (1999).

\bibitem{holographic-beam-splitting-Grier} E. R. Dufresne and D. G.
Grier, {\em Optical tweezer arrays and optical substrates created
with diffractive optics}, Rev. Sci. Instr. {\bf 69}, 1974 (1998);
D.G. Grier and E.R. Dufresne, US Patent 6,055,106, The University
of Chicago (2000).

\bibitem{independent.foci} R. L. Eriksen,  V. R. Daria, and
J. Gl\"uckstad,
{\em Fully dynamic multiple-beam optical tweezers,}
Opt. Express {\bf 10}, 597 (2002); J. E. Curtis, B. A. Koss, and
 D. G. Grier, {\em Dynamic holographic optical tweezers}
Opt. Com. {\bf 207}, 169 (2002); {\em Use of multiple optical
vortices for pumping, mixing and sorting} US Patent Application
20030132373; D. Grier, W. Lopes, and L. Gruber, {\em Configurable
Dynamic Three Dimensional Array}, Intl. Patent Application WO
03/001178 A2 (2003).

\bibitem{peristaltic.foci} B. A. Koss and D. G. Grier,
{\em Optical Peristalsis}, Appl. Phys. Lett. {\bf 82}, 3985
(2003).

\bibitem{fractionation.Grier} K. Ladavac, K. Kasza, and D. G.
Grier, {\em Sorting Mesoscopic Objects with Periodic Potential
Landscapes: Optical Fractionation}, Phys. Rev. E {\bf 70},
010901(R) (2004).

\bibitem{optical.tubetrap} T. Kuga, Y. Torii, N. Shiokawa, T. Hirano,
Y. Shimizu, and H. Sasada, {\em Novel Optical Trap of Atoms with a
Doughnut Beam}, Phys. Rev. Lett. {\bf 78}, 4713 (1997).

\bibitem{optical.tubes} M. J. Renn {\em et al.}, Phys. Rev. Lett. {\bf 75},
3253 (1995); B. T. Wolschrijn, R. A. Cornelussen, R. J. C.
Spreeuw, and H. B. van Linden van den Heuvell, {\em Guiding of
cold atoms by a red-detuned laser beam of moderate power}, New J.
Phys. {\bf 4}, 69 (2002); B. D\'{e}pret, P. Verkerk, D. Hennequin
Opt. Commun. {\bf 211}, 31 (2002).

\bibitem{optical-bottles}
J. Arlt and M. J. Padgett, {\em Generation of a beam with a dark
focus surrounded by regions of higher intensity: the optical
bottle beam}, Opt. Lett. {\bf 25}, 191 (2000); A. Kaplan, N.
Friedman, and N. Davidson, {\em Optimized single-beam dark optical
trap}, J. Opt. Soc. Am. B {\bf 19}, 1233 (2002).

\bibitem{turn.light} D. McGloin, V. Garc\'es-Ch\'avez, and K. Dholakia,
{\em Interfering Bessel beams for optical micromanipulation}, Opt.
Lett. {\bf 28}, 657 (2003); Jennifer E. Curtis and David G. Grier,
{\em Structure of Optical Vortices}, Phys. Rev. Lett. {\bf 90},
133901 (2003).

\bibitem{qudit} G. Molina-Terriza, J. P. Torres, and L. Torner,
{\em Management of the Angular Momentum of Light: Preparation of
Photons in Multidimensional Vector States of Angular Momentum},
Phys. Rev. Lett. {\bf 88}, 013601 (2001); J. Leach, M. J. Padgett,
S. M. Barnett, S. Franke-Arnold, and J. Courtial, {\em Measuring
the Orbital Angular Momentum of a Single Photon, ibid.} {\bf 88},
257901 (2002).

\bibitem{opt.billiard} V. Milner, J. L. Hanssen, W. C.
Campbell, and M. G. Raizen, {\em Optical Billiards for Atoms},
Phys. Rev. Lett. {\bf 86}, 1514 (2001).

\bibitem{superlattice} A. G\"orlitz, T. Kinoshita, T. W. H\"ansch, and A.
Hemmerich, {\em Realization of bichromatic optical superlattices},
Phys. Rev. A {\bf 64}, 011401(R) (2001).

\bibitem{gost} Yu. B. Ovchinnikov, I. Manek, and R. Grimm,
{\em Surface Trap for Cs atoms based on Evanescent-Wave Cooling,}
Phys. Rev. Lett. {\bf 79}, 2225 (1997).

\bibitem{Greiner01} Markus Greiner, Immanuel Bloch, Olaf Mandel,
Theodor W. H\"ansch, and Tilman Esslinger, {\em Exploring Phase
Coherence in a 2D Lattice of Bose-Einstein Condensates },
%
Phys. Rev. Lett. {\bf 87}, 160405 (2001)

\bibitem{accel.lattice} S. R. Wilkinson, C. F. Bharucha, M. C. Fischer,
K.W. Madison, P. R. Morrow, Q. Niu, B. Sundaram, and M. G. Raizen,
{\em Experimental evidence for non-exponential decay in quantum tunnelling},
Nature {\bf 387}, 575 (1997); M. C. Fischer, B.
Guti\'errez-Medina, and M. G. Raizen, {\em Observation of the
Quantum Zeno and Anti-Zeno Effects in an Unstable System}, Phys.
Rev. Lett. {\bf  87}, 040402 (2001).

\bibitem{McGloin03} D. McGloin, G.C. Spalding, H. Melville, W. Sibbet, and
K. Dholakia, {\em Applications of spatial light modulators in atom
optics}, Opt. Exp. {\bf 11}, 158 (2003).

\bibitem{Kuhr01} S. Kuhr, W. Alt, D. Schrader, M. M\"uller, V. Gomer,
and D. Meschede, {\em Deterministic Delivery of a Single Atom}, Science
{\bf 293}, 278 (2001).

\bibitem{Haensel01} D. M. Stamper-Kurn, H.-J. Miesner, A. P. Chikkatur,
S. Inouye, J. Stenger, and W. Ketterle,
{\em Reversible Formation of a Bose-Einstein Condensate},
Phys. Rev. Lett. {\bf 81}, 2194 (1998)
; W. H\"ansel, J. Reichel, P. Hommelhoff,
and T.W. H\"ansch, {\em Magnetic Conveyor Belt for Transporting
and Merging Trapped Atom Clouds},
%
{\em ibid.} {\bf  86}, 608 (2001); E. A. Hinds, C. J. Vale, and M.
G. Boshier, {\em Two-Wire Waveguide and Interferometer for Cold
Atoms}, {\em ibid.} {\bf 86}, 1462 (2001); A. E. Leanhardt, A. P.
Chikkatur, D. Kielpinski, Y. Shin, T. L. Gustavson, W. Ketterle,
and D. E. Pritchard, {\em Propagation of Bose-Einstein Condensates
in a Magnetic Waveguide}, {\em ibid.} {\bf 89}, 040401 (2002).

\bibitem{Orzel} C. Orzel, A. K. Tuchman, M. L. Fenselau, M. Yasuda,
and M. A. Kasevich, {\em Squeezed States in a Bose-Einstein
Condensate}, Science {\bf 291}, 2386 (2001).

\bibitem{patent} O. Steuernagel, {\em Optical Particle Manipulation Systems},
UK-Patent application No 0327649.0, (2003); O. Steuernagel, {\em
submitted}.

\bibitem{CGH} P. C. Mogensen and J. Gl\"uckstad, {\em Dynamic array
generation and pattern formation for optical tweezers}, Opt. Comm.
{\bf 175}, 75 (2000).

\bibitem{cgh.Liesener} J. Liesener, M. Reicherter, T. Haist, and H. J.
Tiziani, {\em Multi-functional optical tweezers using
computer-generated holograms}, Opt. Commun. {\bf 185}, 77 (2000).

\bibitem{Yariv.buch}  A. Yariv, {\em Optical electronics},
(Saunders College Publishing, New York, 1991); A. E. Siegman, {\em
Lasers}, (Oxford Univ. Press, Oxford, 1986).

\bibitem{Haus.buch}  H. A. Haus,
{\em Electromagnetic Noise and Quantum Optical Measurements},
(Springer, Heidelberg, 2000).

\bibitem{Rayleigh.limit} Note, that we cannot implement {\em every} desired
field pattern since the employed monochromatic light only provides
resolution down to the Rayleigh-limit. So, the desired field
pattern has to be sufficiently smooth to be compatible with the
wavelength of the used light. Very high orders in the expansions
of $A_x$ in Equations~(\ref{sum.TEM})
and~(\ref{A.standing.wave.sum.TEM}) cannot be used, they are
incompatible with the paraxiality assumption and therefore do not
allow us to go beyond the Rayleigh-limit.

\bibitem{Tunneling} F. S. Cataliotti, S. Burger, C. Fort, P.
Maddaloni, F. Minardi, A. Trombettoni, A. Smerzi, M. Inguscio,
{\em Josephson Junction Arrays with Bose-Einstein Condensates},
Science {\bf 293}, 843 (2001).

\bibitem{driven.devices} P. H\"anggi, P. Talkner, M. Borkovec,
{\em Reaction-rate theory: fifty years after Kramers}, Rev. Mod.
Phys. {\bf 62}, 251 (1990).

\end{thebibliography}
\end{document}